\documentclass[11pt]{article}
\usepackage{amsmath,amssymb}
\usepackage{amsthm}
\usepackage{mathtools}
\usepackage[noend]{algorithmic}
\usepackage[ruled,vlined]{algorithm2e}
\usepackage{setspace}
\usepackage{url}
\usepackage{makeidx}
\usepackage{enumerate}
\usepackage[top=1in, bottom=1in, left=1in, right=1in]{geometry}
\usepackage{graphicx,float,psfrag,epsfig,caption}
\usepackage[usenames,dvipsnames,svgnames,table]{xcolor}
\definecolor{darkgreen}{rgb}{0.0,0,0.9}
\usepackage[pagebackref,colorlinks=true,pdfpagemode=UseNone,citecolor=OliveGreen,linkcolor=BrickRed,urlcolor=BrickRed,pdfstartview=FitH]{hyperref}
\usepackage{epstopdf}
\usepackage{color}
\usepackage{xr}
\usepackage{subfig}
\usepackage{caption}
\usepackage{graphicx}
\usepackage[utf8]{inputenc}
\usepackage{bm}
\usepackage{cancel}
\usepackage{algorithm2e}

\usepackage{scalerel,stackengine}

\stackMath
\newcommand\reallywidehat[1]{%
\savestack{\tmpbox}{\stretchto{%
  \scaleto{%
    \scalerel*[\widthof{\ensuremath{#1}}]{\kern.1pt\mathchar"0362\kern.1pt}%
    {\rule{0ex}{\textheight}}
  }{\textheight}%
}{2.4ex}}%
\stackon[-6.9pt]{#1}{\tmpbox}%
}
\usepackage[mathscr]{euscript}

\DeclareSymbolFont{rsfs}{U}{rsfs}{m}{n}
\DeclareSymbolFontAlphabet{\mathscrsfs}{rsfs}

\numberwithin{equation}{section}

\newtheoremstyle{myexample} 
    {\topsep}                    
    {\topsep}                    
    {\rm }                   
    {}                           
    {\bf }                   
    {.}                          
    {.5em}                       
    {}  

\newtheoremstyle{myremark} 
    {\topsep}                    
    {\topsep}                    
    {\rm}                        
    {}                           
    {\bf}                        
    {.}                          
    {.5em}                       
    {}  

\theoremstyle{myremark}

\theoremstyle{myremark}

\theoremstyle{myexample}

\definecolor{darkgreen}{rgb}{0.0, 0.5, 0.0}

\newcommand{\bea}{\begin{eqnarray}}
\newcommand{\eea}{\end{eqnarray}}
\newcommand{\<}{\langle}
\renewcommand{\>}{\rangle}
\newcommand{\E}{{\mathbb E}}

\def\fr{\frac}
\def\fr12{\frac{1}{2}}

\def\R{{\mathbb R}}

\def\bR{{\boldsymbol R}}

\def\Unif{{\sf Unif}}

\def\id{{\boldsymbol{I}}}

\def\S{{\mathbb S}}

\def\bpi{\boldsymbol{\pi}}

\def\btheta{{\boldsymbol{\theta}}}
\def\bo{{\boldsymbol{o}}}

\def\bt{{\boldsymbol{\theta}}}

\def\bxi{{\boldsymbol{\xi}}}

\def\bz{{\boldsymbol{z}}}
\def\bx{{\boldsymbol{x}}}

\def\de{{\rm d}}

\def\E{{\mathbb E}}

\def\<{\langle}
\def\>{\rangle}

\def\by{{\boldsymbol{y}}}

\def\b0{{\boldsymbol{0}}}

\def\br{{\boldsymbol r}}

\allowdisplaybreaks
\title{Theory of learning of high-dimensional controlled non-linear dynamical systems (I): models and methods}
\author{Pierfrancesco Urbani\thanks{Université Paris-Saclay, CNRS, CEA, Institut de physique théorique, 91191, Gif-sur-Yvette, France}}
\date{}

\makeindex 
\begin{document}

\hypersetup{linkcolor=RoyalPurple}

\maketitle
\begin{abstract}
Neural ordinary differential equations (neural ODEs) have rapidly gained prominence as a powerful and unifying framework for conceptualizing artificial neural networks, elegantly connecting the continuous-time modeling of dynamical systems with the discrete, data-driven paradigm of modern deep learning. Beyond their practical advantages they offer fresh theoretical insights into the training and generalization properties of neural networks. The distinctive feature of this framework is its dual dynamical nature: inference dynamics, which govern the ODE evolution during forward computation, and training dynamics, which control the optimization of model parameters. This makes neural ODEs a particularly well-suited theoretical framework for studying a large variety of settings such as multi-layer neural networks (ResNets for example), autoregressive models (with next-token generation dynamics), generative models, and recurrent neural networks in theoretical neuroscience. In this work, we introduce a theoretically grounded class of models for studying neural ODEs trained via online stochastic gradient descent. We solve the training dynamics of these models via dynamical mean field theory and derive learning curves in the high-dimensional limit.
\end{abstract}

\tableofcontents

\section{Introduction}
Feed-forward multi-layer perceptrons (MLPs) are the simplest neural network architectures \cite{rumelhart1986learning, hornik1989multilayer}, processing data through sequential layers to produce task-dependent outputs. In doing so, they learn relevant, possibly multiscale, features of the data. Understanding how such feature learning emerges from the high-dimensional, non-linear, out-of-equilibrium training dynamics of model parameters is a central problem in machine learning \cite{bengio2021deep, rawat2017deep}. This question is not only crucial for developing stronger theoretical foundations for neural networks but also represents a fundamental conceptual challenge in science. Indeed, contemporary machine learning practice extends far beyond the boundaries of classical statistical learning theory \cite{vapnik2013nature}. Modern models are heavily overparameterized—so expressive that they can fit pure noise \cite{zhang2016understanding}—yet when trained on meaningful datasets, simple first-order optimization methods successfully guide the weights toward internal representations aligned with the data structure \cite{zhang2016understanding, zhang2021understanding}. In recent years, progress has been made on this problem by both addressing limiting dynamical regimes (online SGD) or linearized models \cite{jacot2018neural, chizat2018global, montanari2023six, mei2018mean, rotskoff2022trainability}. Most recently, \cite{montanari2025dynamical} proposed a solution for non-linear two-layer networks, according to which feature learning and overfitting coexist in overparameterized models but occupy distinct dynamical regimes during training when the model size is large.

Understanding the training dynamics of deeper networks—particularly when trained on complex datasets with multiscale feature structures—remains, however, an open challenge. In this work, we shift the perspective and analyze the problem through the framework of neural differential equations (Neural ODE) \cite{chen2018neural}.
The neural differential equations framework reformulates neural networks as continuous dynamical systems, where discrete layer-wise transformations are replaced by the flow of an ordinary differential equation \cite{chen2018neural, kidger2022neural}. This perspective provides a powerful analytical toolkit: it reveals the underlying dynamics of network evolution, enables the application of dynamical systems theory, and naturally captures multiscale feature learning through the continuous flow. In particular, the framework allows for precise characterization of how features at different scales interact and evolve during training.

Moreover, the Neural ODE perspective establishes explicit connections to other domains where parametric dynamical systems generate controlled outputs to perform specific tasks \cite{chen2018neural, kidger2022neural}. This highly general framework encompasses autoregressive models for sequence generation—including large language models and state-space models  \cite{vaswani2017attention, radford2019language, gu2021efficientS4,gu2020hippo}—flow-based generative models \cite{grathwohl2018ffjord}, and various recurrent architectures, from reservoir computing and echo-state networks \cite{jaeger2004harnessing, maass2002real} to fully recurrent networks used in computational neuroscience \cite{sussillo2009generating}.
In all these frameworks, there are two types of dynamics going on. On one hand there is the dynamics of the dynamical system at hand with its \emph{inference time}, on the other, the parameters of the dynamical systems are learned via a training algorithm whose dynamics is associated to a \emph{training time}.
These dynamics are also associated to different set of variables: for example the inference dynamics pertains to tokens in the context of autoregressive sequence modelling or to the membrane potential of neurons in the case of recurrent neural networks. The training dynamics instead describe the update of the parameters controlling the dynamical system.
It is clear that this two-level dynamical structure poses a serious challenge for theoretical analysis. Studying training dynamics of neural networks in feed-forward context is already per se a complex problem. In the present context this dynamics interferes with the dynamical system itself. The purpose of the present work is to (i) introduce a broad class of models for which this problem can be studied exactly via statistical physics methods and (ii) to develop the theoretical methods to solve them in detail.

We consider a large class of models of Neural ODE whose inference and training dynamics can be solved exactly in the high-dimensional limit. These models are rather idealized but they have some desirable basic features that are shared with much more practical and complex models:
\begin{itemize}
    \item the dynamical systems defining inference dynamics is non-linear in the state variables (tokens, membrane potentials\ldots) and high-dimensional. This dynamics is in general not supposed to sample any target, equilibrium measure and therefore it is generically out-of-equilibrium.
    \item the training dynamics of the parameters of Neural ODE is non-linear and high-dimensional and it is done by optimizing a training objective function either in an online fashion or via empirical risk minimization.
    \item We assume that the dataset where the Neural ODE has to be trained comes from an unknown dynamical system in state space. This dynamics depends on a set of simple features and learning coincides with aligning the model to these features.
\end{itemize}
We will take the perspective of a teacher-student set-up \cite{gardner1989three} in which a teacher dynamical system driven non-linearly by a few feature vectors is learned from a student model, corresponding to a dynamical system whose parameters can be learned via some training scheme (empirical risk minimization or online stochastic gradient descend). The dataset on which the student models is trained is made of the token configurations at time $t=0$ and at some target time $T$ (although we will discuss how this setting can be easily relaxed).

We will then show that the training dynamics of the student model, in the high-dimensional limit, can be studied through dynamical mean field theory. 
As a proof of concept we discuss a simple setting where learning curves can be derived explicitly.

\subsection{Related Literature}
The purpose of the present manuscript is to develop a set of models and methods to study learning high-dimensional controlled dynamical systems.
The corresponding dynamics happens on two levels: first one needs to consider the inference dynamics, when the control variables are fixed. As soon as the dynamical degrees of freedom (such as the tokens) are high-dimensional, understanding inference dynamics requires the study of such high-dimensional dynamics. At the same time learning dynamics changes the control variables in the dynamical system itself. Given that control variables are also high-dimensional, understanding their evolution is itself a problem of dynamics in high dimension.

The study of high-dimensional dynamical systems can be done via dynamical mean field theory (DMFT) \cite{mezard1987spin}.
In recent years non-linear learning dynamics of high-dimensional inference problems \cite{mannelli2019passed, sarao2020marvels, sarao2021analytical,mignacco2021stochasticity,kamali2023stochastic} and feed-forward neural networks  \cite{agoritsas2018out, mignacco2020dynamical, celentano2020estimation,  mignacco2022effective,  bordelon2024dynamical, fan2025dynamical, montanari2025dynamical} has been studied extensively via DMFT.

The generalization to models with both inference and training dynamics is more recent. In the context of recurrent neural networks, this was studied first in \cite{fournier2023statistical} where DMFT was used to analyze the FORCE algorithm \cite{sussillo2009generating} in a simple model of recurrent neural networks. More recently this was extended to models trained via gradient descent \cite{clark2026structure, fournier2026to_appear}. 

Furthermore the problem of controlling dynamical systems can be usually recast into an optimal control setting. In the high-dimensional case where the objective is disordered such framework was studied via statistical physics tools in \cite{urbani2021disordered}. Furthermore when the dynamics is intrinsically high-dimensional but controlled by few degrees of freedom was investigated in \cite{lombardo2021optimization} where the DMFT formalism was developed to study controlled optimization of high-dimensional rugged energy landscapes coming from spin glass theory.

Note that all these frameworks differ from standard low dimensional optimal control settings because in the former cases either the inference dynamics or the learning dynamics or both of them are intrinsically high-dimensional and cannot be mapped to a small (order one) set of controlled ordinary (Markovian) differential equations. This intrinsically low dimensional case has been recently studied in the context of optimizing hyperparameters during training, see for example \cite{mori2025optimal}.

Controlling high-dimensional dynamical systems has a statistical physics counterpart in large deviation theory of out-of-equilibrium systems through macroscopic fluctuation theory \cite{bertini2015macroscopic}. Finally the intrinsically high-dimensional settings that we consider in this work constitutes a classical out-of-equilibrium counterpart of a recent stream of works in quantum optimal control \cite{werschnik2007quantum}.

\subsection{Contribution}
The main contribution of this manuscript is to introduce a set of solvable models where inference and training dynamics can be solved in the intrinsically high-dimensional setting where token/state degrees of freedom and control parameters are high-dimensional variables.
In particular we consider a set of high-dimensional controlled dynamical systems by generalizing those studied in \cite{fournier2023statistical, fournier2025non, fournier2025high}. These dynamical systems are driven by a random Gaussian force field. We introduce additional degrees of freedom in the force fields that play the role of control variables. The statistics of the force fields remains Gaussian but the covariance structure depends now on a set of summary statistics of the control degrees of freedom. 

We consider then a teacher-student setup where the a teacher dynamical system with fixed control variables is used to produce a dataset. This consists in the configuration of the dynamical system at time zero and the configuration after some lag-time $T$. The configuration of the system at this final time depends non-linearly in the control variables even if they appear \emph{linearly} in the force fields.

The teacher model must be fitted by a student dynamical system that has to learn the control variables of the teacher process from the dataset.

We analyze online learning dynamics of the control variables. This is done by embedding the population loss into a Lagrangian scheme where the inference dynamics is enforced by a set of adjoint variables. The dynamics of the adjoint variables runs backward in time at any epoch of training and its high-dimensional. Therefore one can study it via DMFT. The adjoint variables encode the backpropagation of the error signal across time. 

We discuss the analytical structure of the DMFT dynamical system. This reflects the nature of dynamics of the token degrees of freedom, the adjoint variables, and the control parameters. For the class of models that we consider we derive a set of self-consistent equations describing both inference and training time dynamics. The structure of these equations is rather universal but for the class of models that we consider (with Gaussian force fields) one can make analytical progress and reduce the self-consistent stochastic processes to a set of non-linear partial differential equations that capture the dynamics of a set of relevant summary statistics. We show that these equations can be integrated numerically efficiently and compare this analysis with numerical simulations.

\section{The simplest model and the training pipeline}
We consider a setting where a teacher model is represented by a high-dimensional dynamical system governed by a high-dimensional control variable, denoted as $\btheta^*$. The student model is another high-dimensional dynamical system, controlled by a similarly high-dimensional variable $\btheta$. We assume that the student has full knowledge of the teacher model, except for the specific value of the control variable $\btheta^*$. In this context, learning consists in aligning the student’s control vector $\btheta$ with the teacher’s $\btheta^*$. 

\subsection{The data model: the Teacher}
Consider a vector $\bx\in\mathbb{R}^d$ that represents the state of the teacher dynamical system. The dynamics of $\bx$ are autonomous and governed by a vector field $\br : \R^d\times \R^{d_\theta}\to \R^d$, leading to the following evolution equation:

\begin{equation}
    \begin{split}
    \frac{\partial \bx(t,\alpha)}{\partial t} &= -\mu_x(t,\alpha)\bx(t,\alpha) + \br(\bx(t,\alpha),\btheta^*)+\sqrt{\frac{2}{\beta_t}}\bxi_t(t,\alpha) \\
    \bx(0,\alpha)&\sim \Unif(\S_{d-1}(\sqrt d))\:.
    \end{split}
    \label{teacher process}
\end{equation}
The function $\mu_x(t)$ is defined as:
\begin{equation}
    \mu_x(t,\alpha) = \hat \mu(|\bx (t,\alpha)|^2/d)
\end{equation}
and $\hat \mu(z)$ is a rapidly growing function of $z$ for $z\to \infty$.
The noise term $\bxi_t(t,\alpha)$ is white, and its strength is controlled by $\beta_t$: as $\beta_t\to \infty$, the system becomes deterministic apart from the randomness of the initial condition. The vector field $\br$ is assumed to be a Gaussian process, with $\btheta^*\in \S_{d_\theta-1}(\sqrt{d_\theta})$ being a fixed random vector uniformly sampled from the $d_\theta$-dimensional sphere.
The covariance structure of the vector field driving the dynamics is given by:

\begin{equation}
    \E[r_i(\bx,\btheta),r_j(\bx',\btheta')]=\delta_{ij}G\left[\frac{\langle \bx,\bx'\rangle}{d},\frac{\langle \btheta,\btheta'\rangle}{d_\theta}\right]
\end{equation}
where $G$ is, in principle, an arbitrary nonlinear covariance function. Here we will assume that $G(0,x)=0$ for all $x$. This is because in the opposite case, one can always rewrite the vector field $\br$ as the sum of a constant (but random) term, independent on $\bx$ and a centered Gaussian term. 

The process described in Eq.~\eqref{teacher process} is referred to as the \textbf{Teacher Process}. A class of dynamical systems close to Eq.~\eqref{teacher process} has been extensively studied in recent works, including \cite{fournier2023statistical, fournier2025non, fournier2025high, fournier2026chaos}.
In absence of the control variable, the same dynamical system can exhibit fixed point phases and transition to chaotic phases. The maximal Lyapunov exponent can be computed exactly. Finally, when driven by external input in the form of linear, time-dependent, forcing fields, this dynamical system can also show periodic attractors (limit cycles) \cite{fournier2025high}. 

Note that the differential equation in Eq.~\eqref{teacher process} can be also discretized in time. The DMFT analysis for the corresponding discrete time dynamical system can be also extended to this case and has been investigated in \cite{fournier2023statistical,fournier2025high}.

\subsection{The dataset}
To generate a dataset, we simulate the teacher process to produce $n$ independent realizations, each indexed by $\alpha$. While all realizations share the same Gaussian process $\br$ and are driven by the same control parameter $\btheta^*$, they differ in their initial conditions and, possibly the noise realization. The process is simulated over a fixed time window $t \in [0, T_t]$. The resulting dataset is defined as:
\begin{equation}
    \mathcal{D} = \{\bx(0,\alpha), \bx(T_t,\alpha)\}_{\alpha=1,\ldots,n}.
\end{equation}

In the following we will consider this precise setting but we emphasize that the formalism that we develop can be also extended to \emph{monitored} stochastic processes. Consider the Teacher process on a time window $T_t$ and choose $P$ observations points.
A natural dataset is given by
\begin{equation}
    \mathcal{D} = \{\bx(s_j,\alpha)\}_{\alpha=1,\ldots,n}^{j=0,\ldots P}
\end{equation}
such that $s_0=0$ and $s_L=T_t$.
In the limit $P\to \infty$ the observations span the whole stochastic trajectory.
However one can consider the case in which some of these observations are \emph{masked} to mimick masked tokens in trained transformers in natural language processing.

\subsection{Neural ordinary differential equations: the Student}
The dataset generated by the Teacher process is fitted using a Student dynamical system, which has access to all information except the control vector $\btheta^*$. Let $\by(t,\alpha) \in \mathbb{R}^d$ denote the state variable of the Student process at time $t$ and epoch $\alpha$ during training. Its dynamics is governed by the following equation:

\begin{equation}
    \begin{split}
    \frac{\partial \by(t,\alpha)}{\partial t} &= -\mu_y(t,\alpha)\by(t,\alpha) + \br(\by(t,\alpha),\btheta(\alpha)) + \sqrt{\frac{2}{\beta_s}}\bxi_s(t,\alpha), \\
    \by(0,\alpha) &\sim \Unif(\S_{d-1}(\sqrt{d})),
    \end{split}
\end{equation}
where $\mu_y(t,\alpha) = \hat{\mu}(|\by(t,\alpha)|^2/d)$ and $\bxi_s(t,\alpha)$ represents white noise with strength modulated by $\beta_s$.

While the time window $T_t$ and noise strength $\beta_t$ of the Teacher process may differ from those of the Student process ($T_s$ and $\beta_s$, respectively), we focus on the case where $T_t = T_s$ and $\beta_t = \beta_s\to \infty$. In this scenario, learning corresponds to aligning the Student's control parameter $\btheta$ with the Teacher's control vector $\btheta^*$, i.e., $\btheta = \btheta^*$.

\subsection{Training objective}
The training objective at epoch $\alpha$ is defined by the following loss function:
\begin{equation}
    L(\alpha) = \frac{1}{2} \left|\bx(T_s,\alpha) - \by(T_s,\alpha)\right|^2.
    \label{the loss}
\end{equation}
We minimize this loss using online Stochastic Gradient Descent (SGD). Here, $\alpha$ denotes the number of training samples processed, and $\btheta(\alpha)$ represents the configuration of the control variables after $\alpha$ updates (or epochs). Note that the loss function in Eq.~\eqref{the loss} does not include any regularization on the control variable $\btheta$, although this can be easily incorporated if needed.

\subsection{Lagrangian formulation}
The dependence of the loss function $L$ on the control parameter $\btheta$ is fully determined by the endpoint dynamics of the Student model, which is governed by a differential equation controlled by $\btheta$. To explicitly enforce the dependence of $\by$ on $\btheta$, we introduce a Lagrangian formulation of the minimization problem. We define the Lagrangian $\mathcal{L}$ as:
\begin{equation}
\begin{split}
    \mathcal{L} &= L - \int_0^{T_s} \de s \left\langle \bpi(s,\alpha), \dot{\by}(t,\alpha) + \mu_y(t,\alpha)\by(t,\alpha) - \br(\by(t,\alpha),\btheta(\alpha)) \right\rangle \\
    &\quad - \left\langle \bpi(0,\alpha), \by(0,\alpha) - \bx(0,\alpha) \right\rangle,
\end{split}
\end{equation}
where $\bpi(t,\alpha)$ is a Lagrange multiplier introduced to ensure that $\by$ follows the Student process dynamics.

Instead of directly minimizing $L$, we minimize the Lagrangian $\mathcal{L}$. The corresponding Euler-Lagrange equations are:
\begin{equation}
\begin{split}
    &\begin{cases}
        \dot{\by}(t,\alpha) = -\mu_y(t,\alpha)\by(t,\alpha) + \br(\by(t,\alpha),\btheta(\alpha)), \\
        \by(0) = \bx_\alpha(0),
    \end{cases} \\
    &\begin{cases}
        -\dot{\bpi}(t,\alpha) = \nabla_{\by} \left\langle \bpi(t,\alpha), \br(\by(t,\alpha),\btheta(\alpha)-\mu_y(t,\alpha)\by(t,\alpha) \right\rangle, \\
        \bpi(T_s,\alpha) = \nabla_{\by}L.
    \end{cases}
\end{split}
\label{instanton}
\end{equation}
As in general optimal control problems, the state variables evolve forward in time, while the adjoint fields (or momenta in classical mechanics) evolve backward in time.

\subsubsection{A stochastic version of the adjoint field equation as noisy backpropagation}
The adjoint fields $\bpi(t,\alpha)$ can be also interpreted as the backward signal on the weights as obtained from the backpropagation algorithm. This suggests a generalization of the dynamical system in Eq.~\eqref{instanton}. In particular one can consider the case in which the adjoint field equation is noisy 
\begin{equation}
    -\dot{\bpi}(t,\alpha) = \nabla_{\by} \left\langle \bpi(t,\alpha), \br(\by(t,\alpha),\btheta(\alpha)-\mu_y(t,\alpha)\by(t,\alpha) \right\rangle + \sqrt{\frac{2}{\beta_\pi}}\bxi(t,\alpha)
\end{equation}
with $\bxi_\pi(t,\alpha)$ a white noise at inverse temperature $\beta_\pi$. An interesting direction to explore is to investigate the role of this noise in the learning dynamics, especially to try to understand to which point a deformed gradient can be informative and useful during training. This situation may be relevant both for practical applications (with approximate schemes for training \cite{lillicrap2020backpropagation}) and for different settings, mostly coming from computational neuroscience.

\subsection{Online learning}
The minimization of the Lagrangian $\mathcal{L}$ with respect to the control parameter $\btheta$ yields the following update rule:
\begin{equation}
    \begin{split}
    \btheta(\alpha+1) &= \btheta(\alpha) - \eta \nabla_{\btheta} \mathcal{L}, \\
    &= \btheta(\alpha) - \eta \int_0^{T_s} \de s \, \nabla_\btheta \left\langle \bpi(s,\alpha), \br(\by(s,\alpha),\btheta(\alpha)) \right\rangle, \\
    \btheta(0) &\sim \Unif(\S_{d_\theta-1}(\sqrt{d_\theta})),
    \end{split}
\end{equation}
where $\eta$ represents the learning rate.

It is important to note that this formulation does not include any explicit regularization on the control variable $\btheta$. However, the framework can be extended to incorporate regularization terms in the learning dynamics if needed. In particular the DMFT analysis can be extended to study the inclusion of a Ridge regularization term on $\btheta$.

\subsubsection{Online SGD Training Pipeline}
\begin{figure}
    \centering
    \includegraphics[scale=0.3]{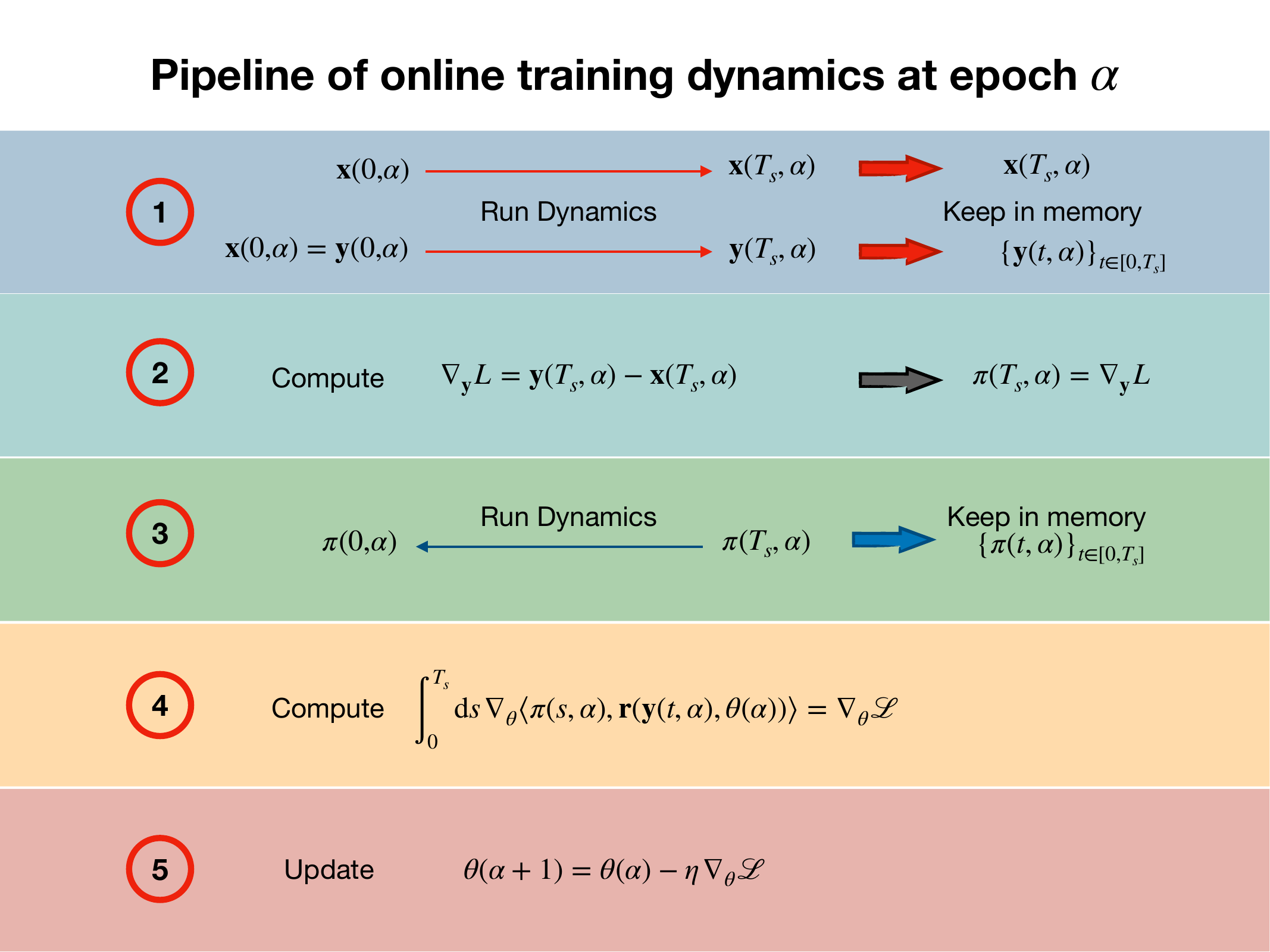}
    \caption{The pipeline of training dynamics.}
    \label{fig:pipeline training}
\end{figure}

The online Stochastic Gradient Descent (SGD) training pipeline is summarized in Fig.\ref{fig:pipeline training}. The dynamics of both $\bx$ and $\by$ are simulated over a time window $T_s$, starting from the same initial condition. The algorithm requires storing the initial and final states of the Teacher process, $\bx(0,\alpha)$ and $\bx(T_s,\alpha)$, which constitute the $\alpha$-th training point, as well as the entire trajectory of the Student process, $\by(t,\alpha)$, for $t \in [0, T_s]$.

Following this, the dynamics of the adjoint fields $\bpi$ are computed backward in time. Once the adjoint trajectory is obtained, the gradient $\nabla_{\btheta} \mathcal{L}$ can be computed, and the control parameter $\btheta$ is updated accordingly.

\subsection{High-Dimensional Asymptotics}
Both the Teacher and Student processes are defined by sets of high-dimensional vectors. We assume that $d \neq d_\theta$ and focus on the high-dimensional limit where $d, d_\theta \to \infty$ with $d_\theta/d \to \sigma$. Importantly, while the dimensionality is sent to infinity, we do not consider time windows that scale with the dimension. In other words, all timescales remain dimension-independent.

\section{Models: generalization}\label{models}
The framework introduced in the previous sections is highly flexible and can be extended in several ways to study several problems. Below, we discuss some generalizations that can be addressed using the same theoretical tools we develop in this manuscript.

\subsection{Mismatched Teacher-Student setting and reasoning}

In the following analysis, we focus on the scenario where the Student has full knowledge of the Teacher's dynamics, except for the feature vector $\btheta^*$. We also assume that $\beta_s, \beta_t \to \infty$, so that both dynamical systems are deterministic.

However, the framework can be easily extended to more general settings:
\begin{itemize}
    \item One can consider $T_s \neq T_t$. For $T_s > T_t$, the Student system is allowed more {\bf reasoning} time, while $T_s < T_t$ imposes a short-answer dynamic. In the case of time-dependent controls, see sec.\ref{dynam_cont}, the situation $T_t<T_s$ is naturally overparametrized.
    \item One can also assume $\beta_s \neq \beta_t$, allowing the two dynamical systems to operate at different noise levels.
\end{itemize}

\subsection{Learning dynamical controls and theory of deep networks}\label{dynam_cont}
In the present manuscript we analyze the case where the control variables are fixed and independent of time. This is true both for the teacher and student dynamical systems.
However this setting can be generalized to cases where the feature vector $\btheta^*$ is time-dependent:
\begin{equation}
    p_\theta \frac{\partial \btheta^*(t)}{\partial t} = \mathbf{f}_{\btheta}(\btheta^*(t)),
    \label{dyn_teacher_theta}
\end{equation}
where $\mathbf{f}_\btheta$ is a vector field and $p_\theta$ is its intrinsic timescale. The student control vector $\btheta(t,\alpha)$ becomes then time and epoch dependent. More importantly, the student has no access to the dynamics in Eq.~\eqref{dyn_teacher_theta} and training coincides with reconstructing the solution of the dynamical system in Eq.~\eqref{dyn_teacher_theta}. The DMFT analysis we detail below can be easily extended to this situation.
The case where the parameters $\bt$ change with time has also resemblance with feed forward neural networks and, in particular, residual neural networks, a perspective that has been put forward in \cite{weinan2018mean}.
Indeed in this case, the configuration of the system at (discrete) time $t$ can be interpreted as the output of the $t$-layer of the network and therefore the control vector $\btheta(t,\alpha)$ can be interpreted as the feature learned at scale $t$ after $\alpha$ epochs.

\subsection{Multilayer Dynamical Systems}

An interesting generalization involves stacking dynamical systems in a multi-layer architecture. Consider a Teacher process indexed by $i = 1, \ldots, L$, with $L$ layers. The dynamics of each layer is given by:
\begin{equation}
    \frac{\partial \bx_i(t,\alpha)}{\partial t} = -\mu_x(t,\alpha,i)\bx_i(t,\alpha) + \br_i(\bx_i(t,\alpha), \bx_{i-1}(t,\alpha), \btheta_i(\alpha)), \quad i = 1, \ldots, L,
\end{equation}
where $\bx_0(t,\alpha)$ can be regarded as an external input. The output is constructed from $\bx_L(T_t,\alpha)$, and a Student process with the same multi-layer structure can be trained accordingly. Each layer is associated with its own feature vector $\btheta_i$, and the vector fields $\br_i$ can be quite general.
The Student dynamics can either follow the same structure as the Teacher process or adopt a mismatched structure.

\subsection{Autoregressive Models and Next-Token Prediction}

Autoregressive models predict future values in a sequence (e.g., words in a sentence) by conditioning on previous values. They are widely used in tasks such as text generation, time series forecasting, and speech synthesis, where order and context are crucial.

In a generic situation, the autoregressive dynamics in tokens space takes the generic form
\begin{equation}
    \bx(t+1) = \bx(t) + \br(\{\bx(s)\}_{s\in[t-\Delta,t]}, \btheta^*)+\bxi(t)
\end{equation}
with $\Delta$ a context length. This dynamical update rule is discrete in time but it can be transformed into a continuous time dynamics with the context length becoming a memory time interval.

In the context of next-token prediction, neural networks are trained to predict a given token in the sequence based on its predecessors. In our current setting, the Teacher dynamical system is re-initialized at each epoch. However, one could consider the case where the initial condition for the next epoch is set to the final state of the previous epoch, i.e., $\bx(0,\alpha+1) = \bx(T_t,\alpha)$. This would correspond to training on a long sequence of measurements from the Teacher dynamical system.

\subsection{Training via Empirical Risk Minimization}
While we primarily consider online SGD for training, practical applications often involve finite training sets, necessitating empirical risk minimization. Although this setting is more challenging to analyze, it can be addressed by extending the formalism we develop here together with the tools discussed in \cite{kamali2023dynamical,kamali2023stochastic,montanari2025dynamical}.

\subsection{Two-layer Teacher}
In the previous sections we cosidered a Teacher process controlled by a single vector $\btheta$ possibly evolving with time.
One simple generalization is a two-layer version of the problem where
\begin{equation}
    \begin{split}
    \frac{\partial \bx(t,\alpha)}{\partial t} &= -\mu_x(t,\alpha)\bx(t,\alpha) + \frac 1{h_t}\sum_{j=1}^{h_t}a^*_j\br_j(\bx(t,\alpha),\btheta^*_j)+\sqrt{\frac{2}{\beta_t}}\bxi_t(t,\alpha) \\
    \bx(0,\alpha)&\sim \Unif(\S_{d-1}(\sqrt d))\:.
    \end{split}
\end{equation}
With $\{\btheta_j^*\}_{j=1,\ldots, h_t}$ a target set of control vectors.
The processes $\br_j$ are Gaussian and can be chosen to be described by the following covariance function
\begin{equation}
    \E[r_{il}(\bx,\btheta)r_{jk}(\by,\tilde \btheta)] = \delta_{ij}\delta_{kl}G\left[\frac 1d \<\bx,\by\>,\frac 1{d_\theta}\<\btheta,\tilde\btheta\>\right]\:.
\end{equation}
The control variables $\{a_j^*\}_{j=1,\ldots,h_t}$ can be made time dependent.
This setting is interesting because one could consider the case in which the $a_j^*$ are broadly distributed leading to a multiscale dynamical process \cite{ren2026emergence}.

\subsection{Two-Layer Student}
Given a two-layer teacher one can consider the case of a two-layer student 
\begin{equation}
    \begin{split}
    \frac{\partial \by(t,\alpha)}{\partial t} &= -\mu_y(t,\alpha)\by(t,\alpha) + \frac 1{h_s}\sum_{j=1}^{h_s}a_j\br_j(\by(t,\alpha),\btheta^*_j)+\sqrt{\frac{2}{\beta_t}}\bxi_s(t,\alpha) \\
    \by(0,\alpha)&\sim \Unif(\S_{d-1}(\sqrt d))\:.
    \end{split}
\end{equation}
where now the control variables are $\{\btheta_j,a_j\}_{j=1,\ldots, h_s}$.

\subsection{Overparametrization}\label{Subsection_over}
It is interesting to consider the case in which $h_t=1$ (single index teacher) and the student dynamical systems has a large number of neurons $h_s\to \infty$. 
If training is performed via empirical risk minimization over a dataset with $n$ points, this problem corresponds to the one studied in \cite{montanari2025dynamical} in a generic supervised learning setting. In particular, in \cite{montanari2025dynamical} it was shown that, at fixed overparametrization ratio (number of training points over the number of parameters in the model), when the student model is trained via empirical risk minimization, training dynamics happens in a two-timescale fashion if the initial complexity of the student is sufficiently small. In particular in the first dynamical regime the student model aligns with the teacher; in a second dynamical regime, happening on timescales scaling with the model size, overfitting kicks in and the model undergoes feature unlearning. {It is clear that we expect such phenomenon happening also in the present setting.}

\subsection{Generative models}
The present setting can be also used to study generative models. Indeed the teacher process maps the uniform distribution on the hypersphere $\S_{d-1}(\sqrt d)$
to a target distribution defined by the teacher process itself.
Therefore the student needs to learn the vector field of the teacher, a setting that corresponds practically to a score-based generative model. 
In an empirical risk minimization context, overfitting the teacher score means producing an associative memory. Conversely, learning the feature vector $\btheta^*$ of the teacher vector field realizes a perfect generative model.
Together with what is discussed in Sec.\ref{Subsection_over}, this implies that the separation of timescales phenomenon discovered in \cite{montanari2025dynamical} can be realized generically also in score-based generative models.

\section{Methods}
To analyze the training dynamics, we need to control the behavior of both the Teacher and Student processes in the high-dimensional limit. To this end, we employ Dynamical Mean Field Theory (DMFT), which allows us to track these dynamics by projecting them onto a few coarse-grained dynamical observables.

We derive the DMFT equations using a path integral representation of the training and inference dynamics \cite{martin1973statistical, janssen1976lagrangean, de1976techniques, de1978dynamics, zinn2021quantum}. While path integral derivations have been previously used to study empirical risk minimization \cite{agoritsas2018out, mignacco2020dynamical, mignacco2022effective, kamali2023stochastic, montanari2025dynamical}, here we apply this approach to study online Stochastic Gradient Descent (SGD). 
We also assume for simplicity that both teacher and student dynamics is noiseless but the inclusion of Langevin noise is immediate.

\subsection{Derivation of the DMFT equations}
The dynamics evolution of the control variables and the tokens of the teacher and student can be rewritten starting from the identity
\begin{equation}
    1=\E_{\bz,\br}\int_{\btheta,\ \bx,\ \by,\ \bpi,\ \hat\bx,\ \hat\by, \hat \bpi}\exp S
    \label{path_int}
\end{equation}
where
\begin{equation}
\begin{split}
    S &= \sum_\alpha \  \langle i \hat  \btheta(\alpha),\  \btheta({\alpha+1}) - \btheta(\alpha) + \eta \int_0^{T_s}\de s \nabla_\btheta \langle \bpi(s,\alpha), \br(\by(t,\alpha),\btheta(\alpha))\rangle\rangle \\
    &+\sum_\alpha \int_s \langle i \hat \bpi(s,\alpha), \dot \bpi(s,\alpha)+ \nabla_{\by}\langle \bpi(s,\alpha),\ \br(\by(s,\alpha),\btheta(\alpha))-\mu_y(s,\alpha)\bpi(s,\alpha)\rangle \rangle\\
    &+\sum_\alpha \int_s \langle i\hat \by(s,\alpha), \dot \by(s,\alpha) +\mu_y(s,\alpha)\by(s,\alpha)-\br(\by(s,\alpha),\btheta(\alpha)) \rangle\\
    &+\sum_\alpha \int_s \langle i \hat \bx(s,\alpha), \dot \bx(s,\alpha) +\mu_y(s,\alpha)\bx(s,\alpha)-\br(\bx(s,\alpha),\btheta^*) \rangle\\
    &+\sum_\alpha \langle i\hat \by(0,\alpha), \by(0,\alpha)-\bz(\alpha)\rangle+\int_\alpha \langle i\hat \bx(0,\alpha), \bx(0,\alpha)-\bz(\alpha)\rangle\\
    &+\sum_\alpha \langle i\hat \bpi(T_s,\alpha), \bpi(\alpha,T_s) - \by(T_s,\alpha)-\bx(T_t,\alpha)\rangle\:.
\end{split}
\label{path_integral}
\end{equation}
The \emph{dynamical action} $S$ is nothing but the Fourier  representation inside the integral in Eq.~\eqref{path_int} of the identities enforcing the dynamical equations  controlling the evolution of all the dynamical variables at play. 
Note that
\begin{equation}
    \bz(\alpha) \sim \mathcal{ N}(0,\id) \ \ \ \ \E[\bz(\alpha),\bz(\beta)]=\delta(\alpha,\beta)\id
\end{equation}
but a different choice can be made.
Given that the path integral in Eq.~\eqref{path_int} equals one, we can average over $\bz(\alpha)$ and over the realization $\br$ \cite{de1978dynamics}.
Consider the average over $\br$.
We can write all the terms appearing in $S$ and containing $\br$ in a rather compact way as
\begin{equation}
    \exp\left[\sum_\alpha\int_s \int_{\tilde \btheta}\int_{\tilde \by} \langle\bR(s,\alpha; \tilde \by,\tilde \btheta), \br (\tilde \by,\tilde \btheta)\rangle\right]
\end{equation}
where $\bR$ is the following operator
\begin{equation}
\begin{split}
    \bR(s,\alpha; \tilde \by,\tilde \btheta) &= \delta(\tilde \by- \by(s,\alpha))\delta(\tilde \btheta- \btheta(\alpha)) \left[ ( \eta\langle i\hat \btheta(\alpha),\nabla_{\tilde\btheta}\rangle +\langle i\hat \bpi(s,\alpha),\nabla_{\tilde \by}\rangle)\bpi(s,\alpha)\right.\\
    &\left.-i\hat \by(s,\alpha) \right]- \delta(\tilde \by- \bx(s,\alpha))\delta(\tilde \btheta- \btheta^*)i\hat \bx(s,\alpha)\:. 
\end{split}
\end{equation}
In this way, taking the average over the random process $\br$ is particularly easy and we get
\begin{equation}
\begin{split}
    &\E \exp\left[\sum_\alpha\int_s \int_{\tilde \btheta}\int_{\tilde \by} \langle\bR(s,\alpha; \tilde \by,\tilde \btheta), \br (\tilde \by,\tilde \btheta)\rangle\right] \\
    &= \exp\left[\frac 12 \sum_{\alpha,\beta}\int_{s,s'} \int_{\tilde \btheta,\tilde \btheta'}\int_{\tilde \by,\tilde \by'}   \langle \bR(s,\alpha; \tilde \by,\tilde \btheta),\bR(s',\beta; \tilde \by',\tilde \btheta') \rangle \right] G\left[\frac{1}{d}\langle \tilde \by,\tilde \by'\rangle,\frac{1}{d_\theta}\langle \tilde \btheta,\tilde \btheta'\rangle\right]\\
    &= \exp\left[-\frac {1}2\sum_{\alpha,\beta}\int_{s,s'} \Omega(\alpha,\beta,s,s') G(q_y,q_\theta) -\frac {1}2 \int_{\alpha,\beta}\int_{s,s'}\langle\hat \bx(s,\alpha),\hat \bx(s',\beta)\rangle G(q_x,1)\right.\\
    & \left.- \sum_{\alpha,\beta}\int_{s,s'} \Gamma(\alpha,\beta,s,s')G(q_{yx},m(\alpha)) - \sum_{\alpha,\beta}\int_{s,s'} \langle \hat \by(s,\alpha),\hat \bx(s',\beta)\rangle G(q_{yx},m(\alpha))\right]
\end{split}
\end{equation}
where
\begin{equation}
\begin{split}
    &\Omega(\alpha,\beta,s,s') =\langle \hat \by(s,\alpha),\hat \by(s',\beta)\rangle \\
    & +\langle\bpi(s,\alpha),\bpi(s',\beta)\rangle \left( \frac{\eta^2\langle\hat \btheta(\alpha),\hat \btheta(\beta)\rangle}{d_\theta}\frac{\partial}{\partial q_\theta} \right.+  \eta^2\left( \frac{\langle \hat \btheta(\alpha), \btheta(\beta)  \rangle}{d_\theta}   \frac{\langle \hat \btheta(\beta), \btheta(\alpha)  \rangle}{d_\theta} \right) \frac{\partial^2}{\partial q_\theta^2}\\
    &+\frac{\langle\hat \bpi(s,\alpha),\hat \bpi(s',\beta)\rangle}{d}\frac{\partial}{\partial q_y}+\left( \frac{\langle \hat \bpi(s,\alpha), \by(s',\beta)  \rangle}{d}   \frac{\langle \hat \bpi(s',\beta), \by(s,\alpha)  \rangle}{d} \right) \frac{\partial^2}{\partial q_y^2}\\
    &+\left.\left( \eta \frac{\langle \hat\btheta(\alpha),\btheta(\beta) \rangle}{d_\theta}\frac{\langle \hat\bpi(s',\beta),\by(s,\alpha) \rangle}{d}+\eta \frac{\langle \hat\btheta(\beta),\btheta(\alpha) \rangle}{d_\theta}\frac{\langle \hat\bpi(s,\alpha),\by(s',\beta) \rangle}{d}\right)\frac{\partial^2}{\partial q_y \partial q_\theta}  \right) \\
    &-\langle \bpi(s,\alpha),\hat \by(s',\beta)\rangle\left(\eta \frac{\langle \hat\btheta(\alpha),\btheta(\beta) \rangle}{d_\theta}\frac{\partial}{\partial q_\theta} + \frac{\langle \hat\bpi(s,\alpha),\by(s',\beta) \rangle}{d}\frac{\partial}{\partial q_y} \ \right)\\
    &-\langle \bpi(s',\beta),\hat \by(s,\alpha)\rangle\left( \eta \frac{\langle \hat\btheta(\beta),\btheta(\alpha) \rangle}{d_\theta}\frac{\partial}{\partial q_\theta} + \frac{\langle \hat\bpi(s',\beta),\by(s,\alpha) \rangle}{d}\frac{\partial}{\partial q_y} \ \right) 
\end{split}
\label{path_integral_final}
\end{equation}
\begin{equation}
    \begin{split}
       \Gamma(\alpha,\beta,s,s') = -\langle \bpi(s,\alpha),\hat \bx(s',\beta)\rangle\left(\eta  \frac{\langle \hat\btheta(\alpha),\btheta^* \rangle}{d_\theta}\frac{\partial}{\partial m} + \frac{\langle \hat\bpi(s,\alpha),\bx(s',\beta) \rangle}{d}\frac{\partial}{\partial q_{yx}} \ \right)\: .
    \end{split}
\end{equation}
The derivatives have to be evaluated in
\begin{equation}
\begin{split}
    q_y&=\frac{1}{d}\langle  \by(s,\alpha), \by(s',\beta)\rangle\ \ \ q_\theta=\frac{1}{d_\theta}\langle  \btheta(\alpha), \btheta(\beta)\rangle\\
    q_x&=\frac{1}{d}\langle  \bx(s,\alpha), \bx(s',\beta)\rangle\\
    q_{yx}&=\frac{1}{d}\langle  \by(s,\alpha), \bx(s',\beta)\rangle\ \ \ m(\alpha)=\frac{1}{d_\theta}\langle  \btheta(\alpha), \btheta^*\rangle
\end{split}
\end{equation}
We introduce the following notation. For any two observables $o,o'\in\{x,\ y,\ \pi\}$ that are time and epoch dependent, we define
\begin{equation}
\begin{split}
    C_{oo'}(t,\alpha,t',\beta)&=\lim_{d\to \infty}\frac1{d}\<\bo(t,\alpha), \bo'(t',\beta)\>\\
    R_{oo'}(t,\alpha,t',\beta)&=\lim_{d\to \infty}\frac 1d \sum_{i=1}^d\frac{\delta o_i(t,\alpha)}{\delta h_{o',i}(t',\beta)}\:.
\end{split}
\label{summary_statistics_inference}
\end{equation}

The first line in Eq.~\eqref{summary_statistics_inference} is the two-time correlation function. The second one defines the response function: it tracks how observable $o$ is changed at time $(t,\alpha)$ when an infinitesimal linear perturbation is added to the equation of motion of the observable $o'$ at time $(t',\beta)$. With this notation we can write
\begin{equation}
    \begin{split}
        q_y &= C_{yy}(t,\alpha,t',\beta)\\
        q_\theta&=C_{\theta\theta}(\alpha,\beta)\\
        q_{yx}&=C_{yx}(t,\alpha,t',\beta)\\
        q_x&=C_{xx}(t,\alpha,t',\beta)\:.
    \end{split}
\end{equation}

Given that the path integral depends only on these summary statistics, one can enforce them by hand and rewrite it as an integral over them. The new dynamical action is proportional to $d$ and therefore in the high-dimensional limit, the path integral can be evaluated via a saddle point.

The corresponding equations can be rewritten in terms of a set of self-consistent stochastic processes for a set of degrees of freedom encoding the dynamics of typical entries of the vectors $\btheta$, $\bpi$, $\bx$ and $\by$. These processes are driven by noise terms and memory kernels whose statistics is has to be computed self-consistenly \cite{mezard1987spin}. These kernels depend on the summary statistics themselves so that one can write
\begin{equation}
\begin{split}
    C_{oo'}(t,\alpha,t',\beta)&=\E[o(t,\alpha)o'(t',\beta)]\\
    R_{oo'}(t,\alpha,t',\beta)&=\E\left[\frac{\delta o(t,\alpha)}{\delta h_{o'}(t',\beta)}\right]
\end{split}
\end{equation}
for carefully chosen stochastic processes.

\subsection{The self-consistent stochastic processes}
We have four self-consistent stochastic processes, for four effective degrees of freedom, namely $\theta$, $x$, $y$ and $\pi$.
Each of these degrees of freedom can be interpreted as the dynamics of an effective (representative) entry of the vectors $\btheta$, $\bx$, $\by$ and $\bpi$.

\subsubsection{Learning dynamics} 
After averaging on the statistics of the fields $\br$ the dynamical action in Eq.~\eqref{path_integral} can be read in terms of a self-consisten discrete-time stochastic proces for an effective control variable $\theta$ evolving as a function of the number of epochs as
\begin{equation}
    \theta(\alpha+1)=\theta(\alpha) + \eta \left[ \sum_{\gamma=1}^\alpha M_R^\theta(\alpha,\gamma)\theta(\gamma) + V(\alpha)\theta^* + \xi_\theta(\alpha)\right]\:.
    \label{DMFT_theta}
\end{equation}
The non-linear interaction between the control degrees of freedom and the inference dynamics of the teacher and student processes are all condensed into the memory kernel $M_R^\theta$, the function $V$ that drives the learning dynamics towards $\btheta^*$ and the noise $\xi_\theta$. 
The latter is a centered Gaussian process with covariance structure given by
\begin{equation}
    \begin{split}
        \E[\xi_\theta(\alpha)\xi_\theta(\beta)] &\equiv M_C^\theta (\alpha,\beta) = \frac 1\sigma \int_0^{T_s}\de s \int_0^{T_s} \de s' \,  C_{\pi\pi}(s,\alpha,s',\beta) \frac{\partial G(q_y,q_\theta)}{\partial q_\theta}
\end{split}        
\label{noise_theta}
\end{equation}
The memory kernel and the function $V$ are instead given by
\begin{equation}
    \begin{split}
        M_R^\theta (\alpha,\beta) & = \frac 1\sigma\int_0^{T_s}\de s \int_0^{T_s}\de s'\left[ C_{\pi\pi}(s,\alpha,s',\beta)\left(\eta R_{\theta\theta}(\alpha,\beta)\frac{\partial^2 G(q_y,q_\theta)}{\partial q_\theta^2}\right. \right.\\
        &\left.\left.- R_{y\pi}(s,\alpha,s',\beta)\frac{\partial^2 G(q_y,q_\theta)}{\partial q_y\partial q_\theta} \right) - R_{\pi y}(s,\alpha,s',\beta)\frac{\partial G[q_{y},q_\theta]}{\partial q_{\theta}}\right] \\
        V(\alpha) &=  -\frac {1}\sigma \int_0^{T_s}\de s \int_0^{T_s}\de s' \sum_{\beta=1}^\alpha R_{\pi x} (s,\alpha,s',\beta) \frac{\partial G(q_{yx},m(\alpha))}{\partial m}\:.
    \end{split}
\label{memory_V_theta}
\end{equation}
The quantities in Eqs.~\eqref{noise_theta} and \eqref{memory_V_theta} depend on a set of dynamical observables ($C_{\pi\pi}$\ldots ) that are summary statistics of the inference dynamics and will be defined in the next section.

The self-consistent process in Eq.~\eqref{DMFT_theta} has a generic structure typical of DMFT \cite{mezard1987spin}. However in the present case, this process is also linear and therefore one can close it by defining the following dynamical observables
\begin{equation}
\begin{split}
C_{\theta\theta}(\alpha,\beta)&=\E[\theta(\alpha)\theta(\beta)]\equiv \lim_{d_\theta\to \infty}\frac 1{d_\theta} \<\btheta(\alpha),\btheta(\beta)\>\\
R_{\theta\theta}(\alpha,\beta)&=\frac 1\eta\E[\frac{\delta \theta(\alpha)}{\delta h_\theta(\beta)}] = \frac 1\eta\lim_{d_\theta\to \infty} \frac 1{d_\theta}\sum_{i=1}^{d_\theta}\frac{\delta\theta_i(\alpha)}{\delta h_\theta^i(\beta)}\\
m(\alpha) &=\E[\theta(\alpha)\theta^*]\equiv \lim_{d_\theta\to \infty}\frac1{d_\theta}\<\btheta(\alpha),\btheta^*\>
\end{split}
\label{Corrs_theta}
\end{equation}
where the averages are over the noise $\xi_\theta(\alpha)$.
The response function $R_{\theta\theta}$ is definied by applying an infinitesimal field $h_\theta$ linearly in the self-consistent dynamics of $\theta$ and looking at how the trajectory of the system changes after this perturbation.
It follows that $R_{\theta\theta}(\alpha,\beta)=0$ is $\alpha\leq \beta$. Finally the function $m(\alpha)$ controls how the control variable at epoch $\alpha$ is aligned with the control variables of the teacher process. 
Note that as soon as $m(\alpha)=1$ and $C_{\theta\theta}(\alpha,\alpha)=1$ the student process has completely aligned with the teacher process and learning stops.

The summary statistics in Eq.~\eqref{Corrs_theta} obey the following discrete-time equations
\begin{equation}
    \begin{split}
        C_{\theta\theta}(\alpha+1,\beta) &= C_{\theta\theta}(\alpha,\beta)+\eta\left[\sum_{\gamma=1}^\alpha M_R^\theta(\alpha,\gamma)C_{\theta\theta}(\beta,\gamma) + V(\alpha)m(\beta)\right.\\
        &\left.+\eta \sum_{\gamma=1}^\beta M_C^\theta (\alpha,\gamma)R_{\theta\theta}(\beta,\gamma) \right]\\
        R_{\theta\theta}(\alpha+1,\beta) &= R_{\theta\theta}(\alpha,\beta)+\delta_{\alpha,\beta}+\eta\sum_{\gamma=\beta}^\alpha M_R^\theta(\alpha,\gamma)R_{\theta\theta}(\gamma,\beta) \\
        m(\alpha+1) &= m(\alpha)+\eta\left[\sum_{\gamma=1}^\alpha M_R^\theta(\alpha,\gamma)m(\gamma) + V(\alpha)\right]\:.
    \end{split}
\end{equation}
Note that we also have
\begin{equation}
    C_{\theta\theta}(\alpha+1,\alpha+1)=C_{\theta\theta}(\alpha,\alpha) + 2 \left(C_{\theta\theta}(\alpha+1,\alpha)-C_{\theta\theta}(\alpha,\alpha)\right) + \eta^2\Delta(\alpha,\alpha)
\end{equation}
with $\Delta(\alpha,\alpha)$ given by
\begin{equation}
    \begin{split}
        \Delta(\alpha,\alpha)&=\sum_{\gamma\gamma'}M_R^\theta(\alpha,\gamma)M_R^\theta(\alpha,\gamma')C_{\theta\theta}(\gamma,\gamma')+M_C^\theta(\alpha,\alpha)+V^2(\alpha) \\
        &+ 2V(\alpha)\sum_\gamma M_R^\theta(\alpha,\gamma)m(\gamma)+2\eta\sum_{\gamma\gamma'}M_R^\theta(\alpha,\gamma)M_C^\theta(\alpha,\gamma')R_{\theta\theta}(\gamma,\gamma')\:.
    \end{split}
\end{equation}
These equations can be integrated forward in training time ($\alpha$) assuming that one knows the correlators of inference time variables (the statistics of $\bx$, $\by$ $\bpi$ during both inference and training time).

\subsubsection{Inference dynamics}
In the previous subsection we showed that the dynamics of the control variables can be descibed in the high-dimensional limit via a self-consistent discrete time stochastic process driven by a set of summary statistics of the token variables and adjoint degrees of freedom. 
In this section we write the self-consistent stochastic processes that describe the dynamics of the tokens of the teacher, student and adjoint variables.

We first introduce the notation
\begin{equation*}
    \partial_1^n\partial_2^mG(z,z')=\frac{\partial^n}{\partial z^n}\frac{\partial^m}{\partial z'^m}G(z,z')\:.
\end{equation*}
Then, the self-consistent stochastic processes for $x,\ y$ and $\pi$ are given by
\begin{equation}
    \begin{split}
        \dot x(t,\alpha) &= -\mu_x(t,\alpha)x(t,\alpha) + \xi_x(t,\alpha)\\
        x(0,\alpha)&=z(\alpha)\\
        \dot y(t,\alpha) &=-\mu_y(t,\alpha)y(t,\alpha) +\sum_{\beta=1}^{\alpha-1}\int_0^{T_s}\de s'  M_R^{y\pi}(t,\alpha,s',\beta)\pi(s',\beta)+\xi_y(t,\alpha)  \\
        y(0,\alpha)&=x(0,\alpha)\\
        -\dot\pi(t,\alpha)&= -\mu_y(t,\alpha)\pi(t,\alpha)+\sum_{\beta=1}^\alpha\int_0^{T_s}\de s M_R^{\pi y}(t,\alpha,s,\beta)y(s,\beta)\\
        &+\sum_{\beta=1}^\alpha\int_0^{T_s}\de s M_R^{\pi x}(t,\alpha,s,\beta)x(s,\beta)+\xi_\pi(t,\alpha)-2 \hat\mu'_y(t,\alpha)C_{y\pi}(t,\alpha,t,\alpha)y(t,\alpha)\\
        \pi(T_s,\alpha)&=y(T_s,\alpha)-x(T_t,\alpha)
    \end{split}
    \label{SCSP}
\end{equation}
where 
$$\mu'_y(t,\alpha) =\left. \frac{\partial \hat \mu_y(z)}{\partial z}\right|_{z=C_{yy}(t,\alpha,t,\alpha)}$$ and 
all noise terms are centered Gaussian processes with the following statistics
\begin{equation}
\begin{split}
    \E[\xi_x(s,\alpha) \xi_x(s',\beta)] &= G(C_{xx}(s,\alpha,s',\beta),1) \equiv M_C^{xx}(s,\alpha,s',\beta)\\
    \E[\xi_y(s,\alpha) \xi_y(s',\beta)] &= G(C_{yy}(s,\alpha,s',\beta),q_\theta)\equiv M_C^{yy}(s,\alpha,s',\beta)\\
    \E[\xi_y(s,\alpha) \xi_x(s',\beta)] &= G(C_{yx}(s,\alpha,s',\beta),m(\alpha))\equiv M_C^{yx}(s,\alpha,s',\beta)\\
    \E[\xi_x(s,\alpha) \xi_y(s',\beta)] &= G(C_{xy}(s,\alpha,s',\beta),m(\beta))\equiv M_C^{xy}(s,\alpha,s',\beta)\\
    \E[\xi_\pi(s,\alpha)\xi_\pi(s',\beta)]&=C_{\pi\pi}(s,\alpha,s'\beta)\partial_1 G(C_{yy}(s,\alpha,s',\beta),q_\theta)\:.
    \end{split}
\end{equation}
Note that if we add a Langevin noise to the dynamics of $\bx$ and $\by$, this translates directly to adding an additional Langevin noise into the corresponding self-consitent stochastic processes. The same remains true when one considers the effect of Langevin noise in the dynamics of $\bpi$.
The memory kernels appearing in Eqs.~\eqref{SCSP} are given by
\begin{equation}
    \begin{split}    M_R^{y\pi}(s,\alpha,s',\beta)&=R_{y\pi}(s,\alpha,s',\beta)\partial_1 G(C_{yy}(s,\alpha,s',\beta),C_{\theta\theta}(\alpha,\beta)) \\
    &- \eta R_{\theta\theta}(\alpha,\beta)\partial_2 G(C_{yy}(s,\alpha,s',\beta),C_{\theta\theta}(\alpha,\beta)) \\
    M_R^{\pi y}(s,\alpha,s',\beta) &= R_{\pi y}(s,\alpha,s',\beta)\partial_1 G(C_{yy}(s,\alpha,s',\beta),C_{\theta\theta}(\alpha,\beta))\\
    &+C_{\pi\pi}(s,\alpha,s',\beta)R_{y\pi}(s,\alpha,s',\beta) \partial^2_1 G(C_{yy}(s,\alpha,s',\beta),C_{\theta\theta}(\alpha,\beta))\\
    &-\eta  R_{\theta\theta}(\alpha,\beta)C_{\pi\pi}(s,\alpha,s',\beta)\partial_1\partial_2 G(C_{yy}(s,\alpha,s',\beta),C_{\theta\theta}(\alpha,\beta))\\
    M_R^{\pi x}(s,\alpha,s',\beta) &= R_{\pi x}(s,\alpha,s',\beta)\partial_1 G(C_{yx}(s,\alpha,s',\beta),m(\alpha))\:.
\end{split}
\end{equation}

\subsection{DMFT for the inference dynamics}
As for the dynamics of the effective control parameter $\theta$, the self-consistent stochastic process for $x$, $y$
and $\pi$ are linear and therefore one obtain a set of partial differential equations for the summary statistics defined in Eq.~\eqref{summary_statistics_inference}.

\subsubsection{The process for $x$}
The dynamics of the $x$ variable is autonomous. We can easily integrate it by closing the self consistent process on correlation and response function.
\begin{equation}
\begin{split}
    \partial_tC_{xx}(t,\alpha,t'\alpha) &= -\mu_x(t,\alpha) C_{xx}(t,\alpha,t',\alpha) + \int_0^{t'}\de s G(C_{xx}(t,\alpha,s,\alpha))R_{xx}(t',\alpha,s,\alpha)\\
    \partial_tR_{xx}(t,\alpha,t'\alpha) &= -\mu_x(t,\alpha) R_{xx}(t,\alpha,t',\alpha) +\delta(t,\alpha,t',\alpha)\\
    C_{xx}(0,\alpha,0,\alpha)&=1
    \end{split}
\end{equation}
From these equations it also follows that
\begin{equation}
    \de_t C_{xx}(t,\alpha,t,\beta)=2\lim_{t'\to t}\partial_tC_{xx}(t,\alpha,t',\beta)\:.
\end{equation}
Note that
\begin{equation}
    C_{xx}(t,\alpha,t',\beta)=\delta_{\alpha\beta}C_{xx}(t,\alpha,t',\alpha)
\end{equation}
since $E[z(\alpha)z(\beta)]=\delta_{\alpha\beta}$.
Together with the assumption that $G(0,x)=0$ we get that
\begin{equation}
    C_{xx}(t,\alpha,t',\beta)=\delta_{\alpha\beta}C_{xx}(t,\alpha,t',\alpha)
\end{equation}
with $C_{xx}(t,\alpha,t',\alpha)$ independent on $\alpha$.
The same is true for $R_{xx}(t,\alpha,t',\beta)=\delta_{\alpha\beta}R_{xx}(t,\alpha,t',\alpha)$.

We now consider the correlation structure between the $x$-process and the $\pi$-process. 
Causality and the independence of the $x$-process from everything else implies that
\begin{equation}
\begin{split}
    C_{x\pi}(t,\alpha,t',\beta)=0\ \ \ \ \forall t,\ t',\ \beta<\alpha\\
    R_{x\pi}(t,\alpha,t',\beta)=0\ \ \ \ \forall t,\ t',\ \beta\leq\alpha\:.
\end{split}
\end{equation}

\subsubsection{The process for $y$}
The student dynamics is summarized by the self-consistent stochastic process for the variable $y$ in Eq.~\eqref{SCSP}. Given that this is linear, we can close it on the relevant summary statistics. We first consider the correlation functions
\begin{equation}
\begin{split}
    &\partial_t C_{yy}(t,\alpha,t',\beta) = -\mu_y(t,\alpha) C_{yy}(t,\alpha,t',\beta) + \sum_{\gamma=1}^{\alpha-1} \int_0^{T_s}\de s' M_R^{y\pi}(t,\alpha,s',\gamma)C_{y\pi}(t',\beta,s',\gamma)\\ 
    &+\sum_{\gamma=1}^{\beta} \int_0^{T_s}\de s'\left[ M_C^{yy}(t,\alpha,s',\gamma)R_{yy}(t',\beta,s',\gamma)+ M_C^{yx}(t,\alpha,s',\gamma)R_{yx}(t',\beta,s',\gamma)\right]
\end{split}
\end{equation}
\begin{equation}
    \de_t C_{yy}(t,\alpha,t,\alpha)=2\lim_{t'\to t}\partial_tC_{yy}(t,\alpha,t',\alpha)
\end{equation}
These equations have to be solved forward in time.
The initial conditions are
\begin{equation}
\begin{split}
    C_{yy}(0,\alpha,0,\alpha) &= C_{xx}(0,\alpha,0,\alpha)\\
    C_{yy}(0,\alpha,t',\beta<\alpha)&=C_{xy}(0,\alpha,t,\beta)=0\:.
\end{split}
\end{equation}
The cross correlations between the $x$-process and the $y$-process are given by
\begin{equation}
\begin{split}
    \partial_t C_{yx}(t,\alpha,t',\beta) &= -\mu_y(t,\alpha) C_{yx}(t,\alpha,t',\beta) + \sum_{\gamma=1}^{\alpha-1} \int_0^{T_s}\de s' M_R^{y\pi}(t,\alpha,s',\gamma)C_{x\pi}(t',\beta,s',\gamma)\\ 
    &+\sum_{\gamma=1}^{\beta} \int_0^{t'}\de s' M_C^{yx}(t,\alpha,s',\gamma)R_{xx}(t',\beta,s',\gamma)
\end{split}
\label{C_yx_eq}
\end{equation}
and this equation must be solved forward in time with the following initial condition
\begin{equation}
\begin{split}
    &C_{yx}(0,\alpha,0,\alpha)=C_{xx}(0,\alpha,0,\alpha)\\
    &C_{yx}(0,\alpha,t',\beta<\alpha)=0 \ \ \ \forall t'\:.
\end{split}
\end{equation}
Note that Eq.~\eqref{C_yx_eq} has a more difficult structure when $\alpha=\beta$. 
In this case we have to add the additional equation
\begin{equation}
    \begin{split}
        \partial_{t}C_{xy}(t,\alpha,t',\alpha) = -\mu_x(t,\alpha) C_{xy}(t,\alpha,t',\alpha) +\sum_{\gamma=1}^\alpha \int_0^{T_s}\de s M_C^{xy}(t,\alpha,s,\gamma)R_{yy}(t',\alpha,s,\gamma)
    \end{split}
\end{equation}
with the diagonal term controlled by the following PDE
\begin{equation}
    \de_t C_{yx}(t,\alpha,t,\alpha) = \lim_{t'\to t}\left(\partial_tC_{yx}(t,\alpha,t',\alpha)+\partial_tC_{yx}(t,\alpha,t',\alpha)\right)\:.
\end{equation}
In other words $C_{xy}$ and $C_{yx}$ must be updated in parallel in any integration scheme when $\alpha=\beta$.

We now come to the equations controlling the response functions.
We have
\begin{equation}
    \begin{split}
        \partial_tR_{yy}(t,\alpha,t',\beta) &= -\mu_y(t,\alpha)R_{yy}(t,\alpha,t',\beta) \\
        &+ \sum_\gamma \int\de s M_R^{y\pi}(t,\alpha,s',\gamma)R_{\pi y}(s',\gamma,t',\beta)+\delta_{\alpha\beta}\delta(t-t')
    \end{split}
\end{equation}
\begin{equation}
    \begin{split}
        \partial_tR_{yx}(t,\alpha,t',\beta) &= -\mu_y(t,\alpha) R_{yx}(t,\alpha,t',\beta)+ \sum_\gamma \int \de s M_R^{\pi y}(t,\alpha,t',\beta)R_{\pi x}(s,\gamma,t',\beta)\:.
    \end{split}
\end{equation}
Given the {\bf resetting dynamics} \cite{evans2011diffusion} on the $y$-process we also have the boundary conditions
\begin{equation}
\begin{split}
    &R_{yy}(0,\alpha,t',\beta)=0\ \ \ \ \forall t', \beta<\alpha\\
    &R_{yx}(0,\alpha,t',\beta)=0\ \ \ \ \forall t', \beta<\alpha\\
    &R_{yx}(t,\alpha,t',\alpha)=0\\
    &R_{xy}=0\:.
\end{split}
\end{equation}

We now examine the correlation between the $y$-process and the $\pi$-process. First, we consider the case where the $y$-process is selected at epoch $\alpha$ and analyze its correlation with the $\pi$-process at earlier epochs $\beta < \alpha$.
\begin{equation}
    \begin{split}
        &\partial_t C_{y\pi}(t,\alpha,t',\beta) = -\mu_y(t,\alpha)C_{y\pi}(t,\alpha,t'\beta) + \sum_{\gamma=1}^{\alpha-1}\int_0^{T_s}\de s M_R^{y\pi}(t,\alpha,s,\gamma)C_{\pi\pi}(s,\gamma,t',\beta)\\
        &C_{y\pi}(0,\alpha,t',\beta)=0\ \ \ \ \forall t',\beta<\alpha
    \end{split}
\end{equation}
At the same time
\begin{equation}
    \begin{split}
        &\partial_tR_{y\pi}(t,\alpha,t',\beta) = -\mu_y(t,\alpha)R_{y\pi}(t,\alpha,t',\beta) + \sum_{\gamma} \int\de s M_R^{y\pi}(t,\alpha,s',\gamma)R_{\pi \pi}(s',\gamma,t',\beta)\\
        &R_{y\pi}(0,\alpha,t',\beta)=0 \ \ \ \ \forall t', \beta<\alpha\:.
    \end{split}
\end{equation}

\subsubsection{The process for $\pi$}
We finally come to the self-consistent stochastic process for the variable $\pi$ representing the effective dynamics of the adjoint fields $\bpi$.
The $\pi$-process runs backward in time as much as the dynamics for $\bpi$. For each datapoint, it is initialized at time $T_s$ and needs to be run backward in inference time.
\begin{equation}
    \begin{split}
        -\partial_tC_{\pi y}(t,\alpha,t',\beta)&=-\mu_y(t,\alpha) C_{\pi y}(t,\alpha,t',\beta) + \sum_\gamma\int \de s M_R^{\pi y}(t,\alpha,s,\gamma)C_{yy}(s,\gamma,t',\beta)\\
        &+\sum_\gamma \int \de s M_R^{\pi x}(t,\alpha,s,\beta)C_{xy}(s,\gamma, t',\beta) - 2\mu'_y(t,\alpha) C_{y\pi}(t,\alpha,t,\alpha)C_{yy}(t,\alpha,t',\beta)\\
        C_{\pi y}(T_s,\alpha,t',\beta) &= C_{yy}(T_s,\alpha,t',\beta)-C_{xy}(T_s,\alpha,t',\beta) \ \ \ \forall t', \beta\leq\alpha
    \end{split}
\end{equation}
\begin{equation}
    \begin{split}
        -\partial_tC_{\pi x}(t,\alpha,t',\beta)&=-\mu_y(t,\alpha) C_{\pi x}(t,\alpha,t',\beta) + \sum_\gamma\int \de s M_R^{\pi y}(t,\alpha,s,\gamma)C_{yx}(s,\gamma,t',\beta)\\
        &+\sum_\gamma \int \de s M_R^{\pi x}(t,\alpha,s,\beta)C_{xx}(s,\gamma, t',\beta)-2\mu'_y(t,\alpha)C_{y\pi}(t,\alpha,t,\alpha)C_{yx}(t,\alpha,t',\beta)\\
        C_{\pi x}(T_s,\alpha,t',\beta) &= C_{yx}(T_s,\alpha,t',\beta)-C_{xx}(T_s,\alpha,t',\beta) \ \ \ \forall t', \beta\leq\alpha\:.
    \end{split}
\end{equation}
Finally we have the self-correlation of the $\pi$-process.
We first consider $\beta<\alpha$ and we get
\begin{equation}
\begin{split}
    -\partial_tC_{\pi \pi}(t,\alpha,t',\beta)&=-\mu_y(t,\alpha) C_{\pi\pi}(t,\alpha,t',\beta) +  \sum_\gamma\int \de s M_R^{\pi y}(t,\alpha,s,\gamma)C_{y\pi}(s,\gamma,t',\beta)\\
    &+\sum_\gamma \int \de s M_R^{\pi x}(t,\alpha,s,\beta)C_{x\pi}(s,\gamma, t',\beta)\\
    &+\sum_{\gamma}\int_0^{T_s} \de s M_C^{\pi\pi}(t,\alpha,s,\gamma) R_{\pi\pi}(t',\beta,s,\gamma) \\
    &-2\mu'_y(t,\alpha)C_{y\pi}(t,\alpha,t,\alpha)C_{y\pi}(t,\alpha,t',\beta)\\
    C_{\pi\pi}(T_s,\alpha,t',\beta) &= C_{yy}(T_s,\alpha,t',\beta) + C_{xx}(T_s,\alpha,t',\beta)\\
    &- C_{yx}(T_s,\alpha,t',\beta)-C_{xy}(T_s,\alpha,t',\beta)\ \ \ \ \forall t', \beta<\alpha\:.
\end{split}
\end{equation}
For $\alpha=\beta$ instead we get
\begin{equation}
\begin{split}
    -\partial_tC_{\pi \pi}(t,\alpha,t',\alpha)&=-\mu_y(t,\alpha) C_{\pi\pi}(t,\alpha,t',\alpha) +  \sum_\gamma\int \de s M_R^{\pi y}(t,\alpha,s,\gamma)C_{y\pi}(s,\gamma,t',\alpha)\\
    &+\sum_\gamma \int \de s M_R^{\pi x}(t,\alpha,s,\alpha)C_{x\pi}(s,\gamma, t',\alpha)\\      
    &+\sum_{\gamma}\int_0^{T_s} \de s M_C^{\pi\pi}(t,\alpha,s,\gamma) R_{\pi\pi}(t',\alpha,s,\gamma) \\
    &-2\mu'_y(t,\alpha)C_{y\pi}(t,\alpha,t,\alpha)C_{y\pi}(t,\alpha,t',\alpha)
\end{split}
\end{equation}    
    with the following boundary conditions
\begin{equation}
\begin{split}    
    C_{\pi\pi}(T_s,\alpha,T_s,\alpha) &= C_{yy}(T_s,\alpha,T_s,\alpha) + C_{xx}(T_s,\alpha,T_s,\alpha)- 2C_{yx}(T_s,\alpha,T_s,\alpha)\:.
\end{split}
\end{equation}
Note that the diagonal-in-time correlation evolves accordingly
\begin{equation}
    -\de_t C_{\pi\pi}(t,\alpha,t,\alpha) = -2\lim_{t'\to t}\partial_t C_{\pi\pi}(t',\alpha,t,\alpha)\:.
\end{equation}
We now turn to response functions.
We have
\begin{equation}
    \begin{split}
        -\partial_t R_{\pi y}(t,\alpha,t',\beta) &= -\mu_y(t,\alpha) R_{\pi y}(t,\alpha,t',\beta) + \sum_\gamma \int \de s M_R^{\pi y }(t,\alpha,s',\gamma) R_{yy}(s,\gamma,t',\beta)\\
        &-2\mu'_y(t,\alpha) C_{y\pi}(t,\alpha,t,\alpha)R_{yy}(t,\alpha,t',\beta)
    \end{split}
\end{equation}
\begin{equation}
    \begin{split}
        -\partial_t R_{\pi x}(t,\alpha,t',\beta)&=-\mu_y(t,\alpha) R_{\pi x}(t,\alpha,t',\beta) + \sum_\gamma\int \de M_R^{\pi y}(t,\alpha,s,\gamma)R_{yx}(s',\gamma,t',\beta)\\
        &- \sum_\gamma\int \de s M_R^{\pi x}(t,\alpha,s,\gamma)R_{xx}(s,\gamma,t',\beta)\\
        &-2\mu'_y(t,\alpha) C_{y\pi}(t,\alpha,t,\alpha)R_{yx}(t,\alpha,t',\beta)\:.
    \end{split}
\end{equation}
The boundary conditions for these equations are 
\begin{equation}
    \begin{split}
        R_{\pi y}(T_s,\alpha,t',\beta) &= R_{yy}(T_s,\alpha,t',\beta)\\
        R_{\pi x}(T_s,\alpha,t',\beta) &= R_{yx}(T_s,\alpha,t',\beta)-R_{xx}(T_s,\alpha,t',\beta)\:.
    \end{split}
\end{equation}
Finally the self-response function of the $\pi$-process is given by
\begin{equation}
    \begin{split}
        -\partial_{t}R_{\pi\pi}(t,\alpha,t',\beta) &= - \mu_y(t,\alpha)R_{\pi\pi}(t,\alpha,t',\beta) )+\delta_{\alpha\beta}\delta(t-t')\\
        &+ \sum_\gamma \int \de s M_R^{\pi y}(t,\alpha,s,\gamma)R_{y\pi}(s,\gamma,t'\beta)\\\
        &-2\mu'_y(t,\alpha) C_{y\pi}(t,\alpha,t,\alpha)R_{y\pi}(t,\alpha,t',\beta)\:.
    \end{split}
\end{equation}
For $\beta<\alpha$ we have the boundary condition
\begin{equation}
    R_{\pi\pi}(T_s,\alpha,t',\beta) = R_{y\pi}(T_s,\alpha,t',\beta)
\end{equation}
while for $\alpha=\beta$ one gets that $R_{\pi\pi}(t,\alpha,t',\alpha)=0$ for all $t'\leq t'$.

\subsubsection{The training objective}
The dynamics of the loss across several epochs is simply given by
\begin{equation}
    \lim_{d\to \infty}\frac 1 dL(\alpha) = \frac{1}{2}\left(C_{yy}(T_s,\alpha,T_s,\alpha) + C_{xx}(T_t,\alpha,T_t,\alpha)- 2C_{yx}(T_s,\alpha, T_t,\alpha)\right)
\end{equation}
Note that as soon as $m(\alpha)=1$ and $C_{\theta\theta}(\alpha,\alpha)=1s$ one obtains that the loss vanishes and the student process has aligned with the teacher one.

\subsection{Global structure of the DMFT equations}
Here we describe what is the global structure of the DMFT equations. 
At each epoch $\alpha$, there are three processes that must be considered: the dynamics of $\bx $, the dynamics of $\by$ and the dynamics of $\bpi$. When this dynamical evolution is resolved, the order parameters entering in the update of $\btheta(\alpha+1)$ can be computed.
However there is a causal hierarchy in the update dynamics of $\bx$, $\by$ and $\bpi$ which gives rise to a sort of backpropagation implementation of the training algorithm. 
This structure is inherited by the DMFT equations and the pseudocode to integrate them is:
\begin{enumerate}
    \item Run in parallel the DMFT equations that involve observables of the type $O_{x\sigma}(t,\alpha,t',\beta)$ or $O_{y,\sigma}(t,\alpha,t',\beta)$ with $\sigma=x,y,\pi$. We assume that $\beta\leq \alpha$.
    \item Run in parallel the DMFT equations that involve observables of the type $O_{\pi\sigma}(t,\alpha,t',\beta)$ with $\sigma=x,y,\pi$. We assume that $\beta<\alpha$.
    \item Update $C_{\theta\theta}(\alpha,\beta)$, $R_{\theta\theta}(\alpha,\beta)$ and $m(\alpha)$.
\end{enumerate}

\section{Learning curves}
In this section we test the formalism developed above in some specific settings. In particular we consider a specific realization of the random driving fields $\br$ and integrate numerically the training dynamics. We then compare these numerical simulations with the numerical solution of the DMFT equations.

\subsection{A specific simple setting}
Consider the specific example in which the teacher process is described by a vector field given by
\begin{equation}
    r_i(\bx,\btheta^*)= \sqrt g_1\sum_{j=1}^{d}\sum_{k=1}^{d_\theta}M_{i;jk}x_j\theta ^*_k +\sqrt g_2\sum_{j<k}^{d_x}T_{i;jk}x_jx_k
    \label{specific process}
\end{equation}
This vector field is linear in $\btheta^*$ but, given that the dynamical system is non-linear, the end point of the dynamics $\bx(T)$ is a non-linear function of the control variables $\btheta^*$.
The tensors $M$ and $T$ are random with the following statistics 
\begin{equation}
    M_{i;jk}=\frac{1}{\sqrt{d\ d_\theta}}Z_{i;jk} \ \ \ \ \ Z_{i;jk}\sim {\cal N}(0,1)
\end{equation}
\begin{equation}
    T_{i;jk}=\frac{1}{d}\tilde Z_{i;jk} \ \ \ \ \ \tilde Z_{i;jk}=\tilde Z_{i;kj}\sim {\cal N}(0,1)\:.
\end{equation}
The covariance structure of $\br$ in Eq.~\eqref{specific process} induced by the statistics of $M$ and $T$ is given by
\begin{equation}
    G(x,y)= g_1xy+\frac{g2}2 x^2\:.
    \label{covariance specific}
\end{equation}
We stress that even if this covariance is linear in $y$ the corresponding dataset generated via the teacher process is a non-linear function of the control parameters $\btheta$.

The form of the vector field in Eq.~\eqref{specific process} is linear in the target control $\btheta$. This makes feature learning possibly happening on timescales of order one. This changes for vector fields $\br$ that are quadratic in $\btheta^*$ inducing a quadratic dependance in $y$ in the covariance function $G$, see Eq.~\eqref{covariance specific}. In this case the learning process posseses a ${\mathbb Z}_2$ symmetry that finite dimensional dynamics has to spontaneously break. This symmetry breaking generates an additional timescale of order $\sim \log d$ before the magnetization $m(\alpha)$ becomes small but of order one. This symmetry breaking phenomenon is ubiquitous in physics. Note that the same phenomenology survives if more than quadratic terms in $\btheta^*$ are added to the control. Indeed even if one adds a more-than-linear odd term in $y$ in $G$, this does not change the symmetry breaking phenomenon given that what makes the timescale of order one is just the linear term.

\begin{figure}[t]
    \centering
    \includegraphics[width=0.49 \columnwidth]{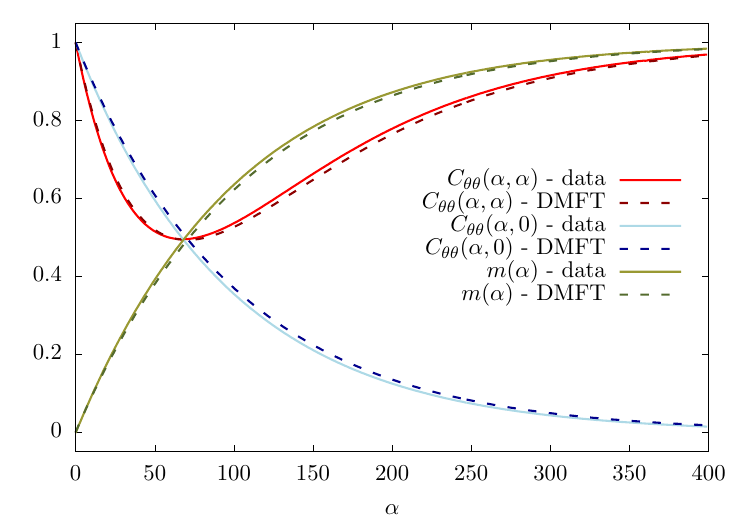}
    \includegraphics[width=0.49 \columnwidth]{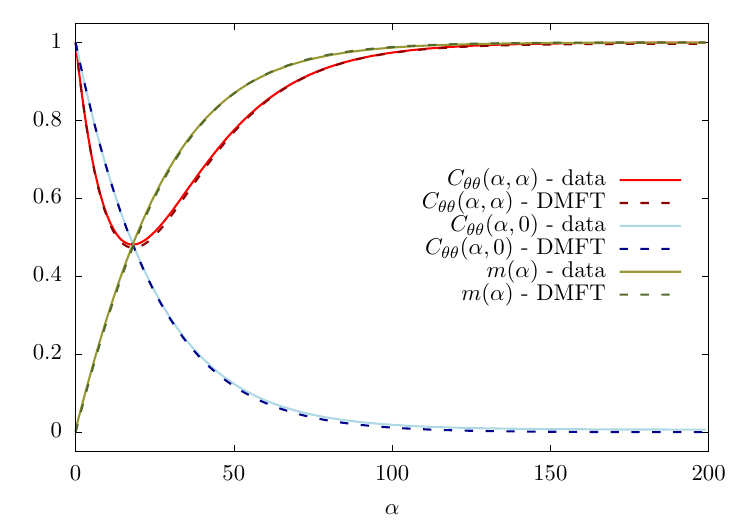}
    \caption{Comparison between DMFT and numerical simulations. We plot tree observables $C_{\theta\theta}(\alpha,\alpha)$, $C_{\theta\theta}(\alpha,0)$ and $m(\alpha)$ as a function of the training epochs. Left panel: Numerical simulations are performed with $d=d_\theta=100$, $g_1=0.8$, $g_2=0.8$, $\mu_x(z)=\mu_y(z)=z$, $\eta=0.1$ and $T_s=0.5$ and $\sigma=1$. Right panel: $\sigma=0.75$ with $d=100$ and $T_s=1$.}
    \label{fig:results_comparison}
\end{figure}

\subsection{Comparison with numerical simulations}
To validate the DMFT analysis we compare the corresponding numerical integration for the summary statistics entering in the DMFT equations with numerical simulations.

Numerical simulations can be performed by integrating explicitly Eq.~\eqref{instanton} with a finite time-step discretization.
In principle, in the high-dimensional limit, the summary statistics dynamics concentrate. However at any finite dimension, different realizations of the training trajectories behave differently and to mitigate this finite dimensional effect we average different trajectories of the relevant summary statistics over several realization of the random driving fields $\br$. 

In practice for each realization of $\br$ we integrate numerically Eq.~\eqref{instanton} and compute the corresponding summary statistics. We then average the corresponding trajectories over different realizations of $\br$.

Our main result is summarized in Fig.~\ref{fig:results_comparison} where we compare the numerical integration of the DMFT equations and numerical simulations. We plot the dynamics of $C_{\theta\theta}$ and $m(\alpha)$ as a function of the epoch number. For the particular control parameters chosen, we have that both $m(\alpha)$ and $C_{\theta\theta}(\alpha,\alpha)$ are approaching one, meaning that the student is aligning with the teacher.

\section{Conclusions and Perspectives}
In this work we initiated the study of coupled inference and training dynamics in neural ordinary
differential equations. We defined a broad class of exactly soluble models where understanding
training performances and inference time behavior can be analyzed exactly, and we developed the
dynamical mean field theory treatment of a simple model within this family. 
This test case allowed us to show that the DMFT analysis
works well to predict learning curves, as evidenced also by comparison with numerical
simulations. In Sec. 3 we discussed generalizations of this framework, to address questions from understanding
training dynamics of deep neural networks, to autoregressive models, to generative models. All
these research directions are very timely and work is in progress to study them with the ideas
and tools developed in this manuscript. In particular two directions are within reach and can be
summarized as follows:
\begin{itemize}
    \item The present work focuses on the online learning setting; however, the methods developed in this work can be extended to study learning via empirical risk minimization. This allows to include memorization effects which are expected to play an important role in generative model, and that are absent when considering online learning. 
    \item It is interesting to generalize the model to a mismatched setting, in which teacher and student differ in  the time window where the inference time runs. In conjunction with an extension of the model to dynamical control variables, this would allow us to study both overparametrization and reasoning effects.
\end{itemize}

We conclude by emphasizing that while the present manuscript is motivated by machine learning and neuroscience applications, the models we develop are sufficiently abstract (and generic)
that can be adapted to other contexts where learning/adaptation is needed. This is the case of controlled matter, system biology (immune system for example) and evolution/adaptation in the
context of ecology and economics. In particular, we believe that  studying the effect of noisy backpropagating error signals is particularly crucial in the biological context.

\paragraph*{Acknowledgements -- }The author warmly thanks Samantha J. Fournier for interesting discussions. Finally the author acknowledges funding by the French government under the France 2030 program (PhOM - Graduate School of Physics) with reference ANR-11-IDEX-0003.  

\bibliography{refs.bib}

\end{document}